\documentclass[fleqn,12pt,twoside]{article}
\usepackage[headings]{espcrc1}

\readRCS
$Id: espcrc1.tex,v 1.2 2004/02/24 11:22:11 spepping Exp $
\usepackage{graphicx}

\newcommand{\AmS}{{\protect\the\textfont2
  A\kern-.1667em\lower.5ex\hbox{M}\kern-.125emS}}

\title{Centrality dependence of heavy flavor production from
single electron measurement in $\sqrt{s_{NN}}$=200 GeV
Au+Au collisions}
 
\author{J.Bielcik\address[MCSD]{Department of Physics, 
	        Yale University, 
 		New Haven, 06511, USA
	} for STAR collaboration\thanks{For the full author list and acknowledgements see Appendix Collaborations in this volume}
        { }
        }
       
\runtitle{Centrality dependence of heavy flavor production from
single electron measurement}
\runauthor{J.Bielcik for the STAR Collaboration}

\begin{document}


\maketitle

\begin{abstract}

We present preliminary measurements of  electron production in p+p, d+Au, and Au+Au collisions at
$\sqrt{s_{NN}}$=200 GeV for transverse momenta 1.5~GeV/$c$~$<p_T<$8~GeV/$c$ as a function of centrality.
These measurements were carried out using the STAR  Time Projection Chamber  and  Barrel
Electromagnetic Calorimeter.  In this manuscript we describe the measurement techniques used to discriminate electrons from hadrons and the method used to evaluate the non-photonic contributions  from semi-leptonic decays of heavy flavor mesons. 
The observed nuclear modification factors, $R_{AA}$, of non-photonic electrons indicate at substantial energy loss of heavy quarks at moderate to high $p_T$.
\end{abstract}

\vspace{0.5cm}

   The measurement of  heavy quark production  helps expand our
knowledge about the nuclear matter produced in  heavy-ion collisions and  furthers our understanding 
of the energy loss mechanism of quarks in the hot and dense medium. 
Due to the large mass of heavy quarks, suppression of small angle gluon 
radiation should reduce their energy loss  and consequently, 
any suppression of  heavy-quark mesons at high $p_T$ is expected to be smaller than
that observed for hadrons consisting of light quarks \cite{Dokshitzer:2001zm,Djordjevic:2005,Armesto:2005}.

  Typically, heavy quark meson production is studied by their reconstruction through hadronic 
decays, e.g. $D^0\rightarrow K^-\pi^+$ \cite{Zhang:2005}. An alternative way to infer infomation about  heavy quark production in heavy ion collisions 
is the study of electrons from semi-leptonic decays of D and B mesons.  This allows STAR to study charm and beauty production 
up to substantially larger $p_T$ than is possible by exploiting the hadronic decay channels. There are several sources of measured electrons. We divide them into
non-photonic electrons (signal) and photonic electrons (background). The non-photonic electrons are mainly 
from semi-leptonic decays of heavy mesons and the Drell-Yan process. The  background photonic electrons are  from $\gamma$ conversions, and  $\pi^0$, $\eta$ Dalitz and light vector meson decays.

   The results presented in this paper were obtained from an analysis of 
data recorded with the STAR detector \cite{Ackermann:2002ad} in years 2003 (p+p, d+Au) 
and  2004 (Au+Au). The two main detector systems used in the analysis were
the Time Projection Chamber (TPC) and the Barrel Electromagnetic Calorimeter (BEMC).
For the data presented here, the BEMC was only half instrumented, limited to the range 
0$<\eta<$1 and 0$<\phi<$2$\pi$, consisting of 2400 towers, each  covering ($\Delta \eta$, $\Delta \phi$)=(0.05, 0.05). A Shower Maximum Detector (SMD)
is located approximately 5 radiation lengths inside each tower module.
It allows the measurement of the shower shape in $\eta$ and 
$\phi$ with high precision ($\Delta \eta$, $\Delta \phi$)=(0.007, 0.007). 
During the 2003 and 2004 runs, STAR used a high tower trigger  (HT), based on the highest energy measured by a single tower in the BEMC, to enrich  the event samples  with high $p_T$ particles that produce signal in the calorimeter, including the electrons. For the Au+Au data, we applied a
 single tower threshold of  $E_T=$~3~GeV/$c$. For the results presented here, aproximately 6.7 M  minimum bias events, 2.6 M high tower trigger events and 4.2 M central events (10$\%$) were used.

\begin{figure}[h!]

\begin{minipage}[t!]{7cm}
\vspace{-0.8cm}
\includegraphics[width=7cm]{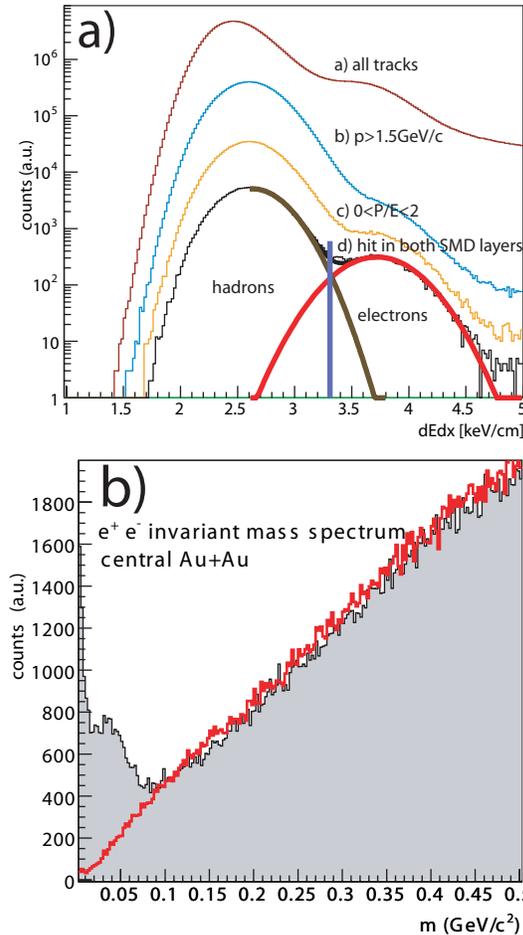}\\

\vspace{-1.7cm}
\caption[dEdx profile after electron selections.]{(a)TPC energy loss of tracks after applications of other electron criteria. (b)Invariant mass of reconstructed e+e- pairs (gray filled area) and combinatorial background (solid red line) in Au+Au central collisions.}
\label{fig:dedx_all}
\end{minipage}

\hspace{\fill}
\begin{minipage}[t!]{8.7cm}
\vspace{-15.7cm}
\hspace{0.5cm}The electron identification was based on a combination of energy loss
in the TPC, energy deposition in the BEMC, and  shower profile in the SMD. We shall refer to electrons in the text meaning both electrons and positrons. Only electron  tracks reconstructed in  the TPC which points to the interaction vertex are taken into account, thus eliminating  a large fraction  of conversion electrons. The measurement of the energy loss, $dE/dx$, in the  TPC  is  a powerful method for particle identification. Hadrons with  $p>$~1.5~GeV/$c$ lose less energy than electrons with the same momentum and can be therefore well separated. 
The resulting  electron candidates  are then extrapolated to the BEMC towers. Their energy deposited in the calorimeter, $E$, is compared with the momentum, $p$, measured 
in the TPC. After traversing the TPC, the electrons deposit their total energy in the BEMC towers, yielding a $p/E$ ratio close to unity, while hadrons showers are less
developed and give $p/E>$1.  Electromagnetic showers are better developed than hadronic showers in 
the SMD detector, yielding to larger reconstructed clusters. Figure~\ref{fig:dedx_all}a shows the energy loss of the tracks after each   electron identification cut. The resulting distribution is fit with two Gaussians to account for
hadrons and electrons. Finally, by applying the $dE/dx$ cut, the electrons can be well separated from hadrons. The fits
to the $dE/dx$ spectra  allow us, in addition, to precisely determine the  remaining 
hadron contamination.

\end{minipage}
\vspace{-1.0cm}
\end{figure}



 \hspace{-0.5cm} A clean sample
of electrons is obtained, with the $p_T$ dependent residual hadron contamination that varies from 
 10 to 15\%. The combined TPC+BEMC hadron discrimination power
is $p_T$ dependent and  of the order of $10^{2}-10^{4}$.

   The data sample for the Au+Au dataset was divided into 3 centrality bins (0-5\%, 10-40\%, and 40-80\% most central).
The electron reconstruction efficiency and acceptance  were determined by embedding simulated
electrons into real events and calculated for each centrality separately. 
For the most central events, the electron reconstruction efficiency 
increases with $p_T$ up to  5~GeV/$c$ and then remains constant at 40\% .

 The dominant sources of photonic electrons are $\gamma$ conversions in the detector material, $\pi^{0}$ and $\eta$ Dalitz decays.
 The contribution from other sources is negligible when compared to systematic uncertainties. The photonic background was evaluated
by identifying both the electron and positron from conversion or Dalitz pair. The identified electrons were combined
with all tracks of opposite charge that passed only loose electron $dE/dx$ cuts to enhance pair reconstruction efficiency.
The invariant mass spectrum of the reconstructed $e^+e^-$ pairs is shown in Fig.~\ref{fig:dedx_all}b. The pairs sample
 contains true backgound pairs as well as
random  combinations (combinatorial background). The combinatorial background 
was estimated from like-sign pairs and is represented on the plot by the solid red line. 

\begin{figure}[h!]

\begin{minipage}[t!]{6.8cm}
\vspace{-1.3cm}
\includegraphics[width=7cm]{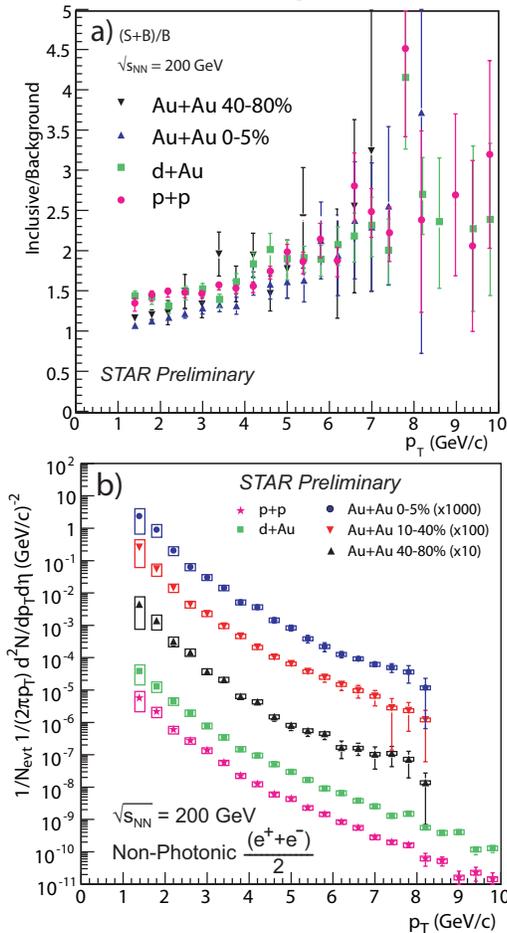}\\

\vspace{-1.7cm}
\caption[Non-Photonic]{(a)Inclusive to photonic electrons ratio as a function of $p_T$ for Au+Au collisions. (b)Background subtracted non-photonic electron spectra for p+p (pink), d+Au (green) and Au+Au collisions with centralities 0-5\% blue, 10-40\% red and 40-80\% black.}
\label{fig:ele}
\end{minipage}

\hspace{\fill}
\begin{minipage}[t!]{9cm}
\vspace{-16.4cm}
\hspace{0.3cm}The pairs from $\gamma$ conversions and $\pi^0$ Dalitz decays
have small invariant masses (see the counts over combinatorial backgound in Fig.~\ref{fig:dedx_all}b) and with a mass cut,
 $m<$140~MeV/$c^2$, most of the photonic background was
removed. The efficiency of the photonic background rejection was determined by embedding $\pi^0$  into real events and is about 60\% for the most central Au+Au events and  slightly decreases with $p_T$. The systematic uncertainties arising
from the treatment of $\eta$ Dalitz decays using $\pi^{0}$ decay kinematics was studied  and found to be negligible for $p_T>$1.5~GeV/$c$.  In Fig.~\ref{fig:ele}a, the ratio of inclusive electrons to photonic background electrons is shown as a function 
of  the $p_T$ of the electrons. For $p_T>$~2.0~GeV/$c$, there is a clear enhancement of electrons with respect to the background. 
This enhancement becomes more evident at higher momentum, where most electrons come from non-photonic sources, such as semi-leptonic decay of heavy quark mesons. Figure~\ref{fig:ele}b shows the preliminary background subtracted non-photonic electron spectra for p+p, d+Au and Au+Au collisions. The error \/ bars \/ are \/statistical \/ and the boxes show the preliminary systematic uncertainties.
 Many NLO pQCD, as well Pythia LO QCD calculations \cite{Ramona:2005}, predict that in a range between 3 and 6 GeV/$c$ the amount of electrons coming from B meson decays become significant. STAR is capable of measuring non-photonic electrons at a momentum range above this transition, making it possible to study in greater detail the interaction between heavy quarks and produced matter.


\end{minipage}
\vspace{-1.0cm}
\end{figure}

\begin{figure}[h!]

\begin{minipage}[t!]{10cm}
\vspace{-1.0cm}
\includegraphics[width=10cm]{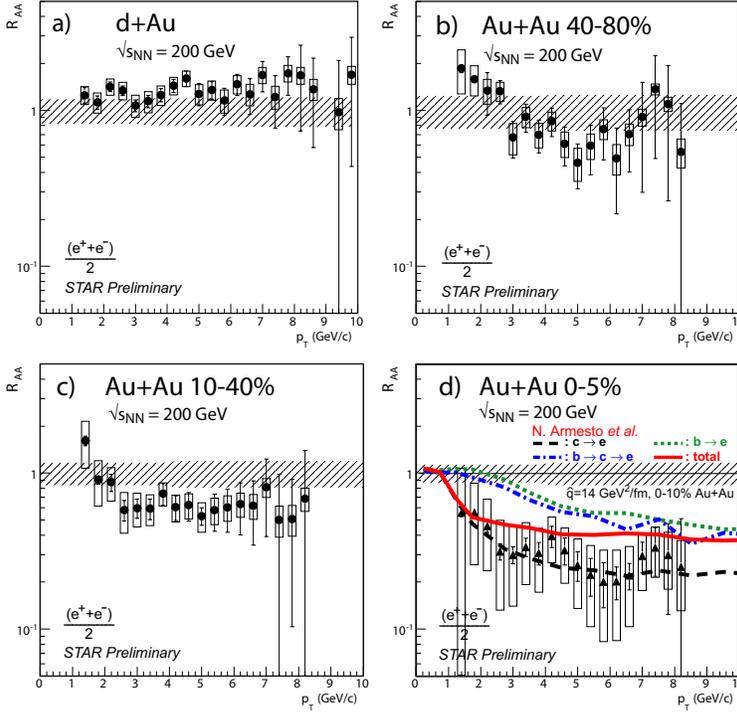}\\

\vspace{-1.5cm}

\caption[Raa]{Nuclear modification factors $R_{AA}$ for (a) d+Au, (b) Au+Au 40-80\%, (c) Au+Au 10-40\%, and 
(d) Au+Au 0-5\% comparison with model \cite{Armesto:2005}.}
 \label{fig:raa}
\end{minipage}

\hspace{\fill}
\begin{minipage}[t!]{5.8cm}
\vspace{-11.2cm}
electrons  indeed come from heavy quark meson decays, this suppression indicates a strong interaction and large energy loss of heavy quarks in the medium created at RHIC. Figure~\ref{fig:raa}d also shows a prediction of the  $R_{AA}$ from a model decribed in \cite{Armesto:2005}, including both the contribution from charm and bottom quarks, with the transport coefficient $\hat q=14~GeV^2/$fm for central Au+Au collisions. Even considering the extreme conditions used in the model, the data points tend to lie below the model prediction at high $p_T$. When compared to the average suppression obtained by identified hadrons at high $p_T$ ($R_{AA}(hadrons)\approx$ 0.2 for central Au+Au), it suggests that the


\end{minipage}
\vspace{-1.0cm}
\end{figure}

Figure~\ref{fig:raa} shows the nuclear modification factors $R_{AA}$ and $R_{dAu}$ for non-photonic electrons, as a function of $p_T$. 
The $R_{dAu}$ ratio (Fig.~\ref{fig:raa}a) is consistent with unity for the entire $p_T$ range, suggesting 
that  non-photonic electron production in d+Au follows a simple binary scaling with respect to p+p collisions.
However, the errors are large and a moderate Cronin type enhancement cannot be ruled out. On the other hand, it is possible to observe an increased suppression from peripheral to central Au+Au events (Fig.~\ref{fig:raa}b-d) with respect to the binary scaling. Supposing that a significant fraction of non-photonic 
  
\hspace{-0.5cm}  amount of energy loss for heavy quarks is comparable to the light quarks, raising questions about the understanding of the energy loss mechanism of partons in dense matter.

      The non-photonic electron spectra measured by the STAR for p+p, d+Au, Au+Au collisions up to $p_T\approx$~8~GeV/$c$ were presented. An increasing suppression of non-photonic electrons with the collision centrality in Au+Au collisions was observed. This may be related to a stronger than predicted interaction between heavy quarks and the medium created at RHIC. The analysis of the full statistics in the 2004 Au+Au run will permit a more detailed study of the medium modifications for heavy quarks allowing a better understanding of quark energy loss mechanisms. The higher statistics data set will also permit initial studies of correlations between non-photonic electrons with charged hadrons that will allow the study of heavy-quark tagged jets in Au+Au collisions at RHIC.


\vspace{-0.2cm} 



\begin{thebibliography}{9}

\bibitem{Dokshitzer:2001zm}
  Y.~L.~Dokshitzer and D.~E.~Kharzeev,
  Phys.\ Lett.\ B {\bf 519}, 199 (2001)
  
\bibitem{Djordjevic:2005}{
	 M.Djordjevic  {\it et al.}, these proccedings}

\bibitem{Armesto:2005}{
	N.~Armesto {\it et al.}, these proccedings}
\bibitem{Zhang:2005}{
         H.~Zhang for STAR collaboration, these proccedings}
\bibitem{Ackermann:2002ad}
  K.~H.~Ackermann {\it et al.},
  Nucl.\ Instrum.\ Meth.\ A {\bf 499}, 624 (2003).

\bibitem{Ramona:2005}{
	M. Cacciari et al., hep-ph/0502203. 
	}



\end{thebibliography}
\end{document}